# A crucial test for astronomical spectrograph calibration with frequency combs

Rafael A. Probst[1, 2]*, Dinko Milaković[3], Borja Toledo-Padrón[4, 5], Gaspare Lo Curto[3],
Gerardo Avila[3], Anna Brucalassi[3], Bruno L. Canto Martins[6], Izan de Castro Leão[6],
Massimiliano Esposito[4, 7], Jonay I. González Hernández[4, 5], Frank Grupp[8, 9],
Theodor W. Hänsch[1], Hanna Kellermann[8, 9], Florian Kerber[3], Olaf Mandel[2],
Antonio Manescau[3], Eszter Pozna[3], Rafael Rebolo[4, 5], José Renan de Medeiros[6],
Tilo Steinmetz[2], Alejandro Suárez Mascareño[4, 10], Thomas Udem[1],
Josefina Urrutia[3], Yuanjie Wu[1, 2], Luca Pasquini[3] and Ronald Holzwarth[1, 2]

**Laser frequency combs (LFCs) are well on their way to becoming the next-generation calibration sources for precision astronomical spectroscopy**[1–6]**. This development is considered key in the hunt for low-mass rocky exoplanets around solar-type stars whose discovery with the radial-velocity method requires cm/s Doppler precision**[7]**. In order to prove such precise calibration with an LFC, it must be compared to another calibrator of at least the same precision. Being the best available spectrograph calibrator, this means comparing it to a second – fully independent – LFC. This test had long been pending, but our installation of two LFCs at the ultra-stable spectrograph HARPS presented the so far unique opportunity for simultaneous calibrations with two separate LFCs. Although limited in time, the test results confirm the 1 cm/s stability that has long been anticipated by the astronomical community.**

First developed for laboratory-based spectroscopy, where they triggered a spectacular gain in accuracy, LFCs have become widely used tools for precision metrology[8]. An LFC is generated by a mode-locked laser that is phase-stabilized to an accurate radio-frequency (RF) reference such as an atomic clock. Its optical spectrum consists of a series of equally spaced, narrow spectral lines (modes), whose frequencies are known to the accuracy of the RF reference. The frequency of the $n^{\text{th}}$ mode is $f_n = f_0 + n\, f_r$, with $f_0$ the offset frequency and $f_r$ the mode spacing. For astronomical echelle spectrographs, this regular pattern of lines, whose frequencies can directly be traced back to the SI second, comes close to an ideal calibrator[9]. LFCs for astronomical applications have thus been developed[1,2,4,10–14,6,15], which stand out through their particularly large mode spacing of > 10 GHz, allowing spectrographs to resolve the mode structure. When monitored on a second spectrograph channel during observations, LFCs permit tracking spectrograph drifts more precisely than ever

[1] Max-Planck-Institut für Quantenoptik, Hans-Kopfermann-Str. 1, 85748 Garching, Germany

[2] Menlo Systems GmbH, Am Klopferspitz 19a, 82152 Martinsried, Germany

[3] European Southern Observatory, Karl-Schwarzschild-Str. 2, 85748 Garching, Germany

[4] Instituto de Astrofísica de Canarias, Vía Láctea s/n, 38200 La Laguna, Tenerife, Spain

[5] Departamento de Astrofísica, Universidad de La Laguna, 38206 La Laguna, Tenerife, Spain

[6] Universidade Federal do Rio Grande do Norte, 59072-970, Natal, RN, Brazil

[7] Thüringer Landessternwarte Tautenburg, Sternwarte 5, 07778 Tautenburg, Germany

[8] Max-Planck-Institut für extraterrestrische Physik, Gießenbachstr. 1, 85748 Garching, Germany

[9] Universitäts-Sternwarte München, Scheinerstr. 1, 81679 München, Germany

[10] Observatoire Astronomique de l'Université de Genève, 1290 Versoix, Switzerland

*r.probst@menlosystems.com

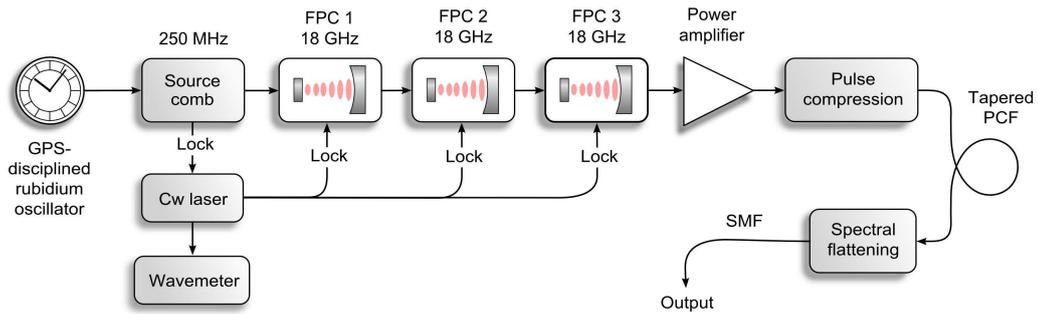

**Figure 1 | Setup of the laser frequency comb (LFC) system.** An Yb-fibre-based LFC with 250 MHz mode spacing is used as a source comb and is stabilized to GPS. A series of three identical Fabry-Pérot cavities (FPCs) increases the mode spacing to 18 GHz (or 25 GHz in the case of LFC2). The FPCs are stabilized in length by a continuous-wave (cw) laser, that itself is stabilized on a transmitted comb mode. Monitoring of the wavelength of this laser with a wavemeter reveals which subset of modes is transmitted through the FPCs. After amplification and pulse compression, the filtered comb spectrum is broadened in a tapered photonic crystal fibre (PCF). Finally, the spectrum is reshaped into a flat-top by the spectral flattening unit. SMF: single-mode fibre.

before[1,2]. LFCs are about to become the future standard calibration sources, replacing thorium-argon arc lamps. The lamps currently limit the attainable precision as their spectrum suffers from strongly irregular line intensity and spacing, blended lines, saturation effects, and from line drifts while the lamp ages.

The resulting gain in precision would greatly benefit exoplanet searches with the radial-velocity method, which looks for periodic Doppler shifts in stellar spectra caused by the gravitational interaction with orbiting planets. This method has enabled the discovery of the first exoplanet around a Sun-like star[16] and continues to yield numerous important discoveries including Earth-mass planets[17,18] and super-Earths[19]. However, the technique has so far been insensitive to Earth-mass planets in the habitable zone of Sun-like stars. An Earth-Sun analogue would manifest in a radial-velocity variation of only ±9 cm/s when viewed edge-on, with a 1-year period. On such time scales, thorium-argon lamps are limited in stability to the m/s level, whereas LFCs are expected to provide a 1 cm/s calibration precision over arbitrary time horizons[9]. LFCs are also anticipated to enable two astronomical tests of fundamental physics: (1) direct measurement of the cosmic expansion rate history, requiring a 1 cm/s precision over about two decades[20]; and (2) measuring the potential cosmological variation in the value of fundamental constants[21].

The HARPS spectrograph[22] is one of several cutting-edge spectrographs that have recently been equipped with an LFC[1,5,10,23–25]. After a number of test campaigns to demonstrate the technique[2,3,26], HARPS has been given a permanently installed LFC (LFC1) as its future routine calibrator in May 2015. The installation was accompanied by the temporary deployment of a second LFC (LFC2), which was thereafter installed at the Wendelstein Observatory for operation with FOCES[23,27]. The scientific goal of the campaign was to characterize the relative performance of two LFCs in a series of repeated simultaneous calibrations in the two HARPS input channels. Similar studies had previously been made with a single LFC for both channels, which indicated a stability of around 2 cm/s[1,2]. Such tests are well suited to disclose potential uncertainties from the spectrograph light injection, imaging system, and image read-out mechanism. Strictly speaking, however, they are incomplete, since they do not rule out any unidentified uncertainties from the LFC itself, which should be common mode in the two channels. This is solved with our simultaneous, relative measurement of two independent LFCs. We therefore report on the most rigorous and precise test conducted so far for proving LFCs as precision calibrators for astronomy.

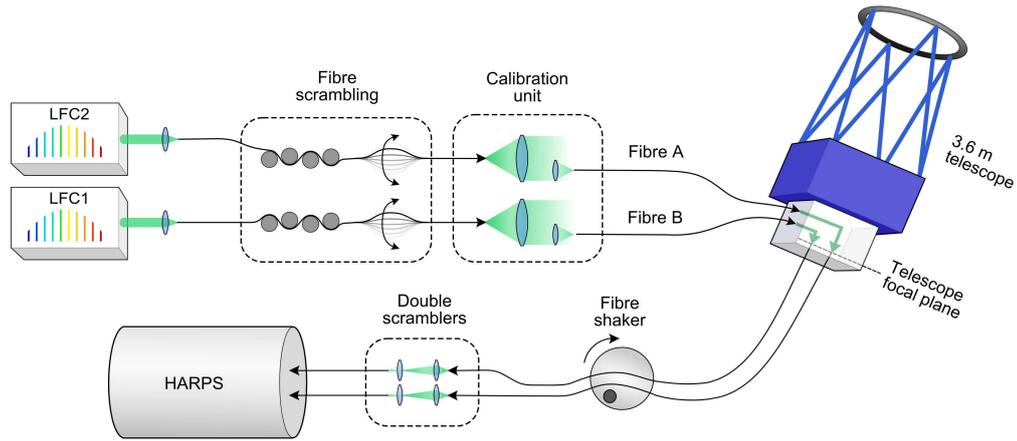

**Figure 2 | Coupling of instruments through optical fibres.** The output of each LFC is injected into a multimode fibre subject to static bends and agitation. In the calibration unit, the fibre output is projected on another set of fibres A and B having motorized fibre inputs and couplers. This allows the operator to select different calibration sources by moving them to different slots. The fibres guide the light to the telescope, where it is projected through the image plane into the fibres leading to HARPS. The last set of fibres is again equipped with a combination of static and dynamic scramblers.

The setup of the LFCs is shown in Figure 1. As a light source, each LFC uses a mode-locked laser with 250 MHz mode spacing and 1040 nm centre wavelength, phase-stabilized to a GPS-disciplined rubidium oscillator as an RF reference. The mode spacing is multiplied to 18 GHz (LFC1) through 3 identical Fabry-Pérot cavities (FPCs) with a finesse of 2600, suppressing all but 1 out of 72 modes. The series of 3 FPCs ensures sufficient suppression for a 1 cm/s calibration accuracy (see Methods). The FPCs are stabilized in length by a continuous-wave laser in a Pound-Drever-Hall scheme[13]. The filtered comb light is then amplified to 12 W of average power, and compressed to a train of ultrashort pulses of 130 fs duration in a grating-prism compressor. This generates sufficient peak power to drive spectral broadening in a tapered photonic crystal fibre[28], which extends the initially infrared spectrum into the visible range. Finally, the broad but structured spectrum is reshaped into a flat-top in the spectral flattening unit[28,29] allowing all comb lines to be of roughly equal signal level on HARPS. LFC2 has a larger mode spacing of 25 GHz, which is matched to the lower resolution of FOCES[23].

Both LFCs are coupled to HARPS through a sequence of multimode fibres (Figure 2). While astronomical applications generally favour multimode fibres over single-mode fibres, as they facilitate efficient throughput of light from astronomical sources, they also come with the issue of modal noise[30]: the beam profile at the output of a multimode fibre depends on light injection conditions at its input and on fibre bend. This is particularly true for coherent light as emitted by an LFC, which acquires laser speckles through modal interference[26]. As a countermeasure, we agitate the first and last set of fibres with electric motors. This makes the speckle pattern change quickly, while the spectrograph averages over it with its much longer exposures. In addition, we have a static bend structure in place on the first fibre pair in the sequence, as well as a double scrambler[31] on the last pair. These static scramblers globally homogenize the beam by coupling different spatial fibre modes to one another.

Figure 3a shows a part of the echellogram recorded with HARPS using the two LFCs. The full echellogram is included in the supplementary information. From the two-dimensional image, we extract one-dimensional spectra by projecting each echelle order of each channel on the spectral direction (Figure 3b). For spectrograph calibration, we determine the line centres by fitting each line with a Gaussian function (see Figure 3c). A coarse thorium-argon calibration is used for mode identification to unambiguously assign frequencies $f_n$. Figure 3d shows the line amplitude spectrum, adjusted by the grating blaze function. The flat-top region ranges from 455 to 691 nm (76 % of the HARPS spectral range), and is flat within 13 % (root-mean-square).

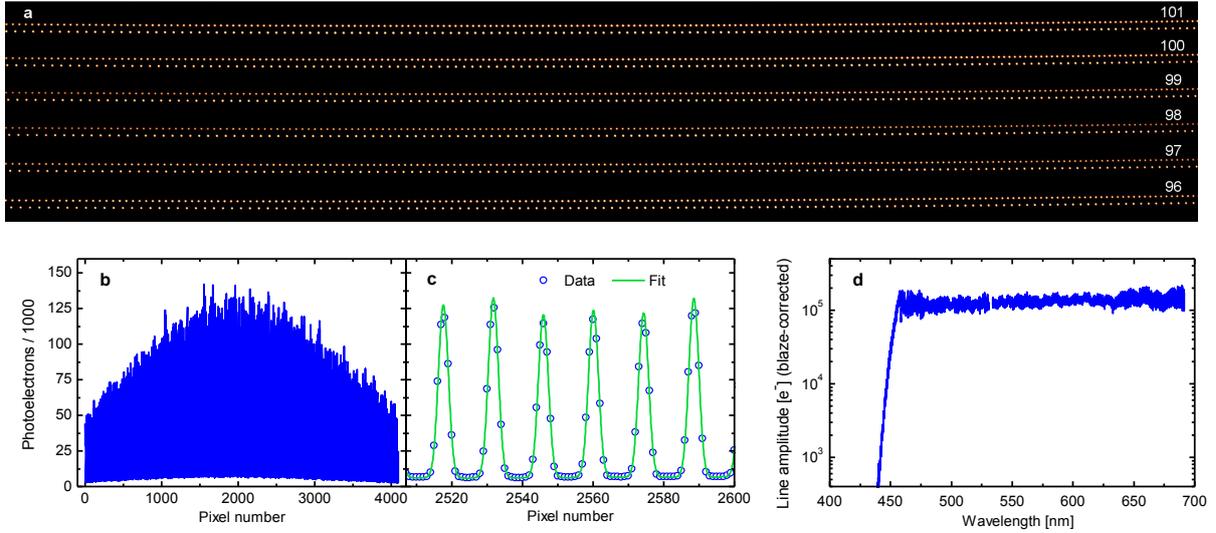

**Figure 3 | Two laser frequency combs (LFCs) on the HARPS spectrograph. a**, Part of the echellogram with the echelle orders labelled with their physical diffraction order. The upper part of each order is channel A (here using LFC1, 18 GHz mode spacing), and the lower part is channel B (LFC2, 25 GHz mode spacing). **b**, Channel A, order 99 (centre wavelength: 618 nm) after data extraction. **c**, Section of the data in part b with a Gaussian function fitted to each line. **d**, Fitted peak values of the LFC lines, corrected by the blaze function of each echelle order.

To assess the relative stability, we repeatedly calibrate the spectrograph and measure by how much each exposure is shifted relative to a reference exposure at the beginning of the series (Figure 4a). This tracks spectrograph drifts, which are generally assumed to be equal in both channels, within the limits given by photon noise. With LFC1 in channel A and LFC2 in channel B, the standard deviation in the differential shifts is of 4.4 cm/s with a photon noise of 3.2 cm/s. We attribute the deviation from photon noise to residual modal noise, as the fiber scrambling configuration had to be well optimized to reduce excess noise to this level (see supplementary information). Interestingly, the remaining uncertainty appears purely statistical in nature: if we bin subsequent exposures and repeat the analysis for the series of binned exposures, we see the stability improving along with the photon noise (Figure 4b). This continues down to a standard deviation of about 1 cm/s, where the limited size of our statistical sample inhibits us from demonstrating a further improvement.

Besides their relative stability, we also compared the two LFCs for their absolute consistency. For this we changed the calibration source from LFC2 to LFC1 on one channel, while continuously keeping LFC2 on the other channel to safely track spectrograph drifts. This reveals a 49 cm/s systematic shift between the calibrations from different LFCs (Figure 5). This was validated using a second, independent analysis (see supplementary Sect 4). The discrepancy can be explained by the differing illumination of the fibres A and B in the calibration unit and/or variation in the intensity of the LFC light. When altering the alignment at this point, we observed comparable shifts despite the use of mode scramblers on the subsequent fibres (see supplementary Section 3).

The LFC as a calibrator is not only extremely stable and reproducible, it also features unparalleled accuracy. We demonstrate how to make use of this property by verifying the absolute velocity of a solar system body. For this we select the dwarf planet Ceres, whose light features the reflected, Doppler shifted solar spectrum. The spectrograph was calibrated with LFC1 (channel A) and LFC2 (channel B) before the observation. While Ceres was observed on channel A, LFC2 remained on channel B to adjust the calibration for spectrograph drifts (Figure 6). By cross-correlating the observed, calibrated spectrum with a mask modelling solar spectrum, we measure a Doppler shift of -21 800.6 m/s. From JPL Horizons[32] we expect this value to be -21 797.6 m/s, thus showing our ability to measure absolute Doppler velocities with about 3 m/s accuracy. At this level, our measurement is limited by solar activity and by inhomogeneity of the albedo of Ceres in combination with its rotation[33]. Observing sunlight reflected from solar system bodies offers the benefit of evenly averaging the solar spectrum across the entire solar disc. In combination with the accuracy provided by the LFC, this opens up new possibilities such as an accurate measurement of the solar gravitational redshift by comparing various spectral lines formed in different altitudes of the Sun to laboratory wavelengths.

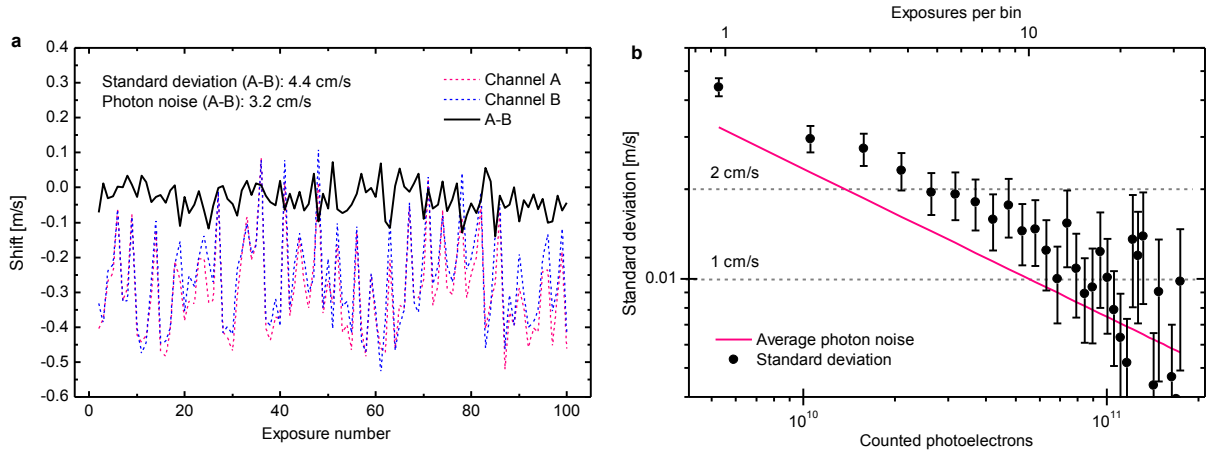

**Figure 4 | Relative stability measurement of two laser frequency combs (LFCs).** LFC1 (18 GHz mode spacing) is in channel A and LFC2 (25 GHz mode spacing) in channel B. **a**, Series of 100 spectrograph calibrations with one exposure every 61 s (integration time: 30 s, readout time: 22.6 s), total duration of the series: 102 min. **b**, Results obtained with binned exposures of increasing size. The filled circles represent the standard deviation in A-B. The error bars quantify the uncertainty of the standard deviation estimated from the size of the statistical sample.

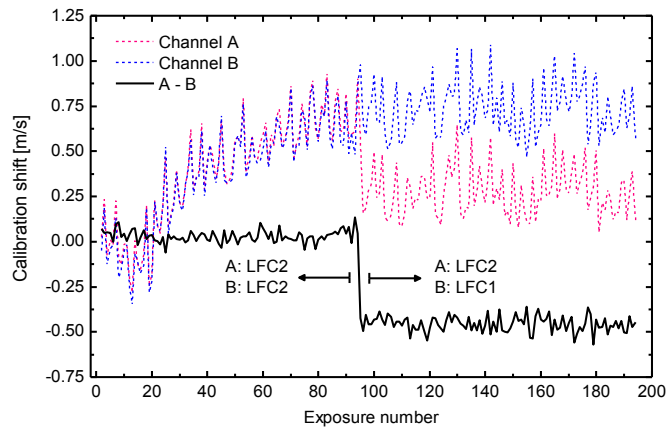

**Figure 5 | Calibration reproducibility with two different LFCs: mean shift in line positions.** After exposure number 94 of 194, the source in channel B is changed from LFC2 to LFC1. LFC2 is continually kept on channel A to track spectrograph drifts. One acquisition was taken every 61 s (integration time: 30 s. Readout time: 22.6 s), with a 100 min pause before the change to the other LFC. The analysis was performed by comparing mean shifts of individual lines. Changing the source in channel B leads to an average shift of 49 cm/s, after taking into account the different line structure of LFC1 (see Methods). We ascribe this shift to differing spectrograph illuminations (see text), not to the LFCs themselves.

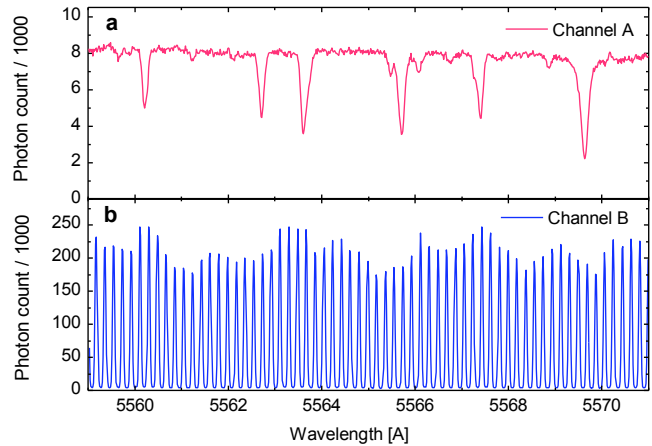

**Figure 6 | Observation of Ceres. a** Ceres spectrum observed in channel A, order 110, integrated over 900 s. The wavelength scale has been calibrated with the LFC at the start of the night and was continually adjusted for spectrograph drifts using the LFC that was kept on Channel B. **b** 25 GHz LFC spectrum that was observed in channel B at the time of the Ceres observation.

In this paper we have shown two LFCs with a relative stability of 1 cm/s over 102 minutes. Since the two LFCs are fully independent, each LFC by itself must be stable to this level. Common to both LFCs is merely the GPS reference, which features proven stability. The most demanding astrophysical applications require the level of stability demonstrated in this paper, but maintained over years or decades. This calls for a comparison of two LFCs with a very long time base. Our present work is an important step into this direction by probing the limiting statistical and systematic effects. Our results show the critical influence of reproducible fibre illumination conditions and mitigation of modal noise even more clearly than earlier investigations[2,26]. For finding Earth-Sun analogues, it will also be essential to develop more advanced methods for separating weak orbital signals from the noise created by stellar activity phenomena. Precise, LFC-calibrated observations are invaluable in this respect[34].

**Acknowledgements**

We cordially thank the technical staff of the La Silla Observatory for their assistance during the installation and the test run. We are equally grateful to Prof. Philip St. J. Russell and his group from the Max Planck Institute for the Science of Light for their support on the development of tapered photonic crystal fibres to broaden the LFC spectrum.

**Author contributions**

R.H., F.G., L.P. and G.L.C. initiated the project. G.L.C., T.S., H.K., G.A., Y.W., A.S.M, F.K., R.A.P., A.M., B.L.C.M., I.d.C.L., L.P. O.M., E.P, J.R.d.M., R.H. and J.U. were at the La Silla Observatory for on-site activities during the campaign. T.S., Y.W., H.K., A.B. and F.G. provided and operated the LFCs. O.M., E.P. and J.U. programmed the software to integrate the LFCs into the infrastructure of the observatory. G.L.C., B.L.C.M. and F.K. operated the spectrograph. A.M. and G.A. designed, built and optimized the light injection and scrambling in multimode fibres. R.A.P., D.M., G.L.C., A.S.M. and I.d.C.L. analysed the LFC data. B.T.P and J.I.G.H. analysed the Ceres observation. R.H., Th.U., T.W.H., R.R. and J.R.d.M. supervised the work. R.A.P. wrote the manuscript. All authors discussed and commented on the manuscript.

## Methods

**Operation and characterization of the LFCs.** During the measurements presented above, the LFCs were operated at a repetition rate (offset frequency) of 18 GHz (5.27 GHz) and 25 GHz (9.70 GHz), respectively. Both values can be altered within certain limits, which we did for some measurements in the supplementary material. This is indicated where being the case. For spectral broadening, the LFCs employed a tapered photonic crystal fibre (PCF) with a design very similar to the one described in Refs. 5, 35. The main difference is the higher air-filling fraction (80 %), which allows the PCF to generate a somewhat wider spectrum. The broad but structured spectrum was flattened out using a liquid-crystal-on-silicon spatial light modulator (SLM) as an adaptive spectral filter[13,28,29]. The filter was set to truncate the spectrum at −20 dB below its peak within the visible range. This is accomplished with the help of a small CCD spectrometer that is integrated in the spectral flattening unit. However, when the spectrum appears flat on the system's internal spectrometer, it is not measured to be flat on HARPS, due to differences in spectral sensitivity and wavelength-dependent losses in the feed-through. Therefore, we first recorded a flattened but uncorrected spectrum with HARPS. From this we derived a spectral correction factor, which could then be taken into account by the software controlling the flattening unit. This yielded the flat-top spectrum shown in Figure 3d. The procedure was carried out separately with each LFC.

The accuracy provided by the spectral lines of the LFC can be influenced by the weak presence of unwanted modes (side-modes) that are not fully suppressed by the Fabry-Pérot cavities (FPCs) in the setup. With the side-modes being unresolved by HARPS, they can shift the centroid of the calibration lines, in case their intensities are asymmetric around the observed line centre. Using the method described in Ref. [13], we measure the finesse of our FPCs to be 2588 (geometric average for the FPCs in LFC2). From this measurement, we calculate the suppression of the strongest side-mode after the series of three FPCs to be 111.4 dB for LFC1 and 102.8 dB for LFC2. The subsequent spectral broadening is known to re-amplify side-modes. An upper limit on the side-mode amplification can be gained from the width of the broadened spectrum[36], yielding 62 dB with our PCFs. This method has been shown to be reliable in several measurements[36] including a PCF that is very similar to ours[35]. The remaining side-mode suppression should ensure a calibration accurate to 0.2 cm/s for LFC1 and 1 cm/s for LFC2. The centroid of a mode can also be shifted through a distortion of its shape while being transmitted through an FPC, in case the filter function is not accurately centred on the mode. For a series of three FPCs with the measured finesse, we calculate the worst-case line shift to be 0.5 cm/s for LFC1 and 0.3 cm/s for LFC2. The dominant error is thus 0.5 cm/s for LFC1 and of 1 cm/s for LFC2.

Both LFCs were connected to the same RF reference, which was a Datum Model 9390 atomic clock with a 10 MHz GPS-disciplined Rb oscillator, which was part of the existing infrastructure of the observatory. The unit had previously been characterized relative to another 10 MHz Rb oscillator (Standford Research Systems PRS10, not locked to GPS) over 24 hours using a frequency counter. Employing a 30 s integration time in this comparison (identical to the integration time we used in our HARPS exposures), the two RF signals proved to be stable within $5.6 \times 10^{-12}$. This corresponds to a radial velocity uncertainty of <0.2 cm/s.

**Light delivery to HARPS.** Both LFCs are each coupled into a 1 mm thick acrylic plastic fibre with a 980 μm core and a numerical aperture (NA) of 0.51. We impose a static bend structure on this fibre, which couples the fibre spatial modes to one another. This globally homogenizes the far-field beam profile at the fibre output, thereby making it less dependent on the illumination conditions at the input, which reduces the related systematic calibration uncertainties. However, for a highly coherent light source such as an LFC, such static scramblers are not sufficient. Interference between different spatial fibre modes creates a speckle pattern at the fibre output, which changes with the slightest motion of the fibre, inducing calibration errors on the m/s scale[26]. Therefore, we attach a small electric motor to the fibre. The motor spins an eccentric weight to make the fibre vibrate. This causes the speckle pattern to change quickly, making the light behave similar to spatially incoherent light on the much longer time scales of the spectrograph exposures. The large NA and core diameter of the plastic fibre makes this process very efficient. This is because it supports a large number of spatial modes, and thus creates a large number of very small speckles, that are highly sensitive to the motions of the fibre. We have also tested a silica fibre with a 200 µm octagonal core in the place of the plastic fibre. The octagonal core by itself is an excellent static mode scrambler, as it is very effective in mixing spatial fibre modes. Yet, the calibration results with this fibre were consistently a factor of 4–5 above the photon noise limit. Most likely, this is due to its smaller core and lower NA, causing the fibre to support a lower number of spatial modes, which reduces the effect of the fibre agitation.

The two scrambled plastic fibres, that each carry the light of an LFC, lead to the calibration unit (see Figure 2), where their output produces a large spot of light, into which the motorized inputs of the fibres A and/or B can be moved to choose a calibration source for each channel. The fibres A and B are a set of silica fibres with a round core of 300 µm in diameter. This pair of fibres leads to the telescope, where the light is projected through the image plane onto the entrance facets of the next set of fibres that feed the spectrograph. From this point, the calibration light takes the same path that star light from the telescope takes when used for spectroscopy. The last set of fibres has a 70 µm circular core. They incorporate a double scrambler[31] to redistribute the spatial modes within the fibre, which is an excellent static mode scrambler. The purpose of the double scrambler is to mitigate the effect of telescope guiding errors on spectroscopy of astronomical objects. This is needed because guiding errors entail varying illumination of the fibre entrance. Although our calibration tests also profit from this static type of scrambler, we additionally installed a dynamic scrambler that agitates the last set of fibres. It consists of a rotating wheel with an off-centred support moving the fibres up and down. While testing the relative stability of the two LFCs, we could only come close to the photon-noise limit with this fibre shaker added to the last set of fibres (see supplementary information).

**Data acquisition and processing.** The HARPS data were typically recorded in sequences of spectrograph exposures with one exposure per minute. For the data shown in the main article we used a 30 s integration time and a 22.6 s readout time. The images are first processed with the HARPS pipeline which automatically subtracts a dark image and the detector bias, performs spectral localization, flat-fielding, cosmic ray removal and spectral extraction  (optimal extraction after Horne[37]). The extracted spectrum contains 4096 data points per echelle order in every channel (see Figure 3b for an example of a single extracted echelle order). The centre positions of the comb lines are determined by fitting them with a model function. Experimenting with several line models, we found that a simple Gaussian fit leads to unreliable results. The reason for this is the relatively strong continuum background in the LFC spectrum, that keeps the signal from dropping to zero between the calibration lines (see Figure 3b, c, and supplementary material). The background can have a non-zero slope that – if not properly taken into account by the data analysis – can shift the detected line positions. Remarkably, the background level consistently follows the structure of the envelope of the comb lines. Tests that we conducted with strongly structured spectra revealed global calibration errors of up to 1 m/s using a simple Gaussian fit. Hence, we add a linear polynomial to the fit, modelling each line as a Gaussian function and the background around each line as a first-order polynomial:

$$f(x) = a_0 + a_1\,x + a_2\,\exp\left[-\frac{(x - a_3)^2}{a_4^{\,2}}\right]$$

Here, $x$ is the position on the sensor in pixels. We define $a_3$ as the centre position of the line. This function is fitted to the data using chi-square minimization with the Levenberg-Marquardt method. The uncertainties of the data points in the fit are given by $\sqrt{|N| + R^2}$, where $N$ is the number of photons detected in each pixel, and $R$ is the readout noise. Besides the fitted parameters, the fit routine also returns the uncertainties of the parameters computed through Gaussian error propagation from the uncertainties of the data points, which are usually dominated by photon noise.

The primary interest of our data analysis lies in a precise assessment of the relative shifts in the calibration. To this end, we determine the overall shift in each channel for each exposure relative to a reference exposure. The reference exposure is consistently chosen to be the first exposure (exposure number 1) of each analysed sequence of exposures, unless stated otherwise. Our standard way of measuring this shift is to average the individual shifts of all lines, weighting each line by its inverse variance in $a_3$ as returned by the fit routine. This yields traces for channel A (red dashed line) and channel B (blue dashed line) as seen in Figure 4 and Figure 5. We remove spectrograph drifts by taking the difference between the two traces (A–B, black solid line), thus revealing the relative stability of calibrations in the two channels. We use the standard deviation of A–B as a measure of stability and compare it to the associated photon noise (compound uncertainty in $a_3$ over all lines). Note that switching channel B to carry light from LFC1 after exposure 94 in Figure 5 requires interpolation between the comb lines in order to make the calculation of line shifts meaningful. We do this by linearly interpolating between the lines of LFC2 in the reference exposure to derive the expected line positions of LFC1 and their uncertainties.

**Ceres observation.** Ceres was observed over 66 minutes on April 18, 2015. The observation was made in four separate spectrograph exposures, each integrated over 900 s. The photometric centres of the exposures are: 08:33:13, 08:50:31, 09:06:59 and 09:24:04 UTC. The Ceres spectra were cross-correlated with a spectral mask based on a list of solar lines at laboratory wavelengths with 3625 lines within the spectral range of the LFC. In the actual solar spectrum, these wavelengths are shifted by effects such as the convective blue shift and the gravitational redshift. Empirical use of this mask has shown that this shifts the derived radial velocities by 99.5 m/s (average over multiple observations[33]), which we subtract from the Doppler shifts that we measure. The solar spectrum is calibrated by creating a wavelength solution from the known optical frequencies of the LFC lines versus their observed positions on the detector. This is constructed as a piecewise 3$^{rd}$ order polynomial across each master block of 512 pixels in width. This allows us to take into account stitching errors from the manufacturing process of the CCD[3,26]. The calibration is then adjusted by the spectrograph drift as seen with the LFC on channel B during the observation of Ceres. After cross-correlating the calibrated Ceres spectrum with the mask, the cross-correlation function is fitted with a Gaussian function, whose centre indicates the average shift of the lines relative to their positions in the mask. With this we determine the Doppler shift to be -21 800.6 m/s on average over the four exposures (individual values: -21 831.1, -21 812.1, -21 791.2 and -21 767.8 m/s). From the known orbit of Ceres relative to the Sun, and from the known motion of the observer relative to Ceres, we compute the predicted Doppler shift to be -21 797.6 m/s on average (individual values: -21 829.0, -21 808.2, -21 787.2 and -21 765.8 m/s).

## Additional references

*Supplementary Information*

# A crucial test for astronomical spectrograph calibration with frequency combs

Rafael A. Probst[1, 2]*, Dinko Milaković[3], Borja Toledo-Padrón[4, 5], Gaspare Lo Curto[3], Gerardo Avila[3], Anna Brucalassi[3], Bruno L. Canto Martins[6], Izan de Castro Leão[6], Massimiliano Esposito[4, 7], Jonay I. González Hernández[4, 5], Frank Grupp[8, 9], Theodor W. Hänsch[1], Hanna Kellermann[8, 9], Florian Kerber[3], Olaf Mandel[2], Antonio Manescau[3], Eszter Pozna[3], Rafael Rebolo[4, 5], José Renan de Medeiros[6], Tilo Steinmetz[2], Alejandro Suárez Mascareño[5, 10], Thomas Udem[1], Josefina Urrutia[3], Yuanjie Wu[1, 2], Luca Pasquini[3] and Ronald Holzwarth[1, 2]

[1] Max-Planck-Institut für Quantenoptik, Hans-Kopfermann-Str. 1, 85748 Garching, Germany
[2] Menlo Systems GmbH, Am Klopferspitz 19a, 82152 Martinsried, Germany
[3] European Southern Observatory, Karl-Schwarzschild-Str. 2, 85748 Garching, Germany
[4] Instituto de Astrofísica de Canarias, Vía Láctea s/n, 38200 La Laguna, Tenerife, Spain
[5] Departamento de Astrofísica, Universidad de La Laguna, 38206 La Laguna, Tenerife, Spain
[6] Universidade Federal do Rio Grande do Norte, 59072-970, Natal, RN, Brazil
[7] Thüringer Landessternwarte Tautenburg, Sternwarte 5, 07778 Tautenburg, Germany
[8] Max-Planck-Institut für extraterrestrische Physik, Gießenbachstr. 1, 85748 Garching, Germany
[9] Universitäts-Sternwarte München, Scheinerstr. 1, 81679 München, Germany
[10] Observatoire Astronomique de l'Université de Genève, 1290 Versoix, Switzerland

*r.probst@menlosystems.com

## 1 Instruments

### 1.1 The HARPS spectrograph

The High Accuracy Radial Velocity Planet Searcher (HARPS) is a fibre-fed, cross-dispersed echelle spectrograph with two channels[1,2]. It is located at the La Silla Observatory in the Chilean Atacama desert and is operated by the European Southern Observatory (ESO). The spectrograph is contained in a vacuum vessel with a pressure of below 0.01 mbar and a temperature stability of 10 mK at its interior, which minimizes spectrograph drifts. HARPS is equipped with one object and one reference fibre for simultaneous calibration, currently using thorium-argon lamps as standard calibrators. It covers a wavelength range of 380–690 nm with 72 echelle orders and a resolution of 115 000. Its camera consists of a mosaic of two CCDs, each with a format of 4096 × 2048 px². HARPS is scientifically devoted to the search for extrasolar planets via radial-velocity measurements. Through its excellent stability, HARPS has become the most successful planet hunter of its kind.

HARPS has played a key role in introducing laser frequency combs (LFCs) for astronomical applications. In a collaboration comprising ESO, the Max Planck Institute of Quantum Optics (MPQ), and Menlo Systems GmbH, a total of five test campaigns have been conducted on HARPS to demonstrate spectrograph calibration with an LFC. The campaigns were carried out in January 2009[3], March 2010, November 2010[4], January 2011[4], and February 2012. The LFCs used in these tests were developed at MPQ and Menlo Systems, while similar systems were also developed by other groups and tested at other spectrographs[5–10]. The test campaigns on HARPS yielded several ground-breaking results, such as a new and improved atlas of solar lines[11] and the first planetary orbit measured with an LFC[4]. Absolute calibration of HARPS was demonstrated with previously unparalleled accuracy, which allowed accounting for structures in the CCD pixelation that previously went unnoticed[3,11]. Further, a previously unmatched calibration repeatability of 2.5 cm/s was demonstrated[4]. Very similar results for the repeatability were later also attained with LFCs at other facilities[5,12]. LFCs at spectrographs other than HARPS have scientifically been applied in observations of the "Sun as a star"[12,13]. Such observations are used to search ways to discern stellar activity from the radial-velocity signature of low-mass planets. In another scientific study, an LFC was utilized to create a new atlas of uranium-neon lines[14].

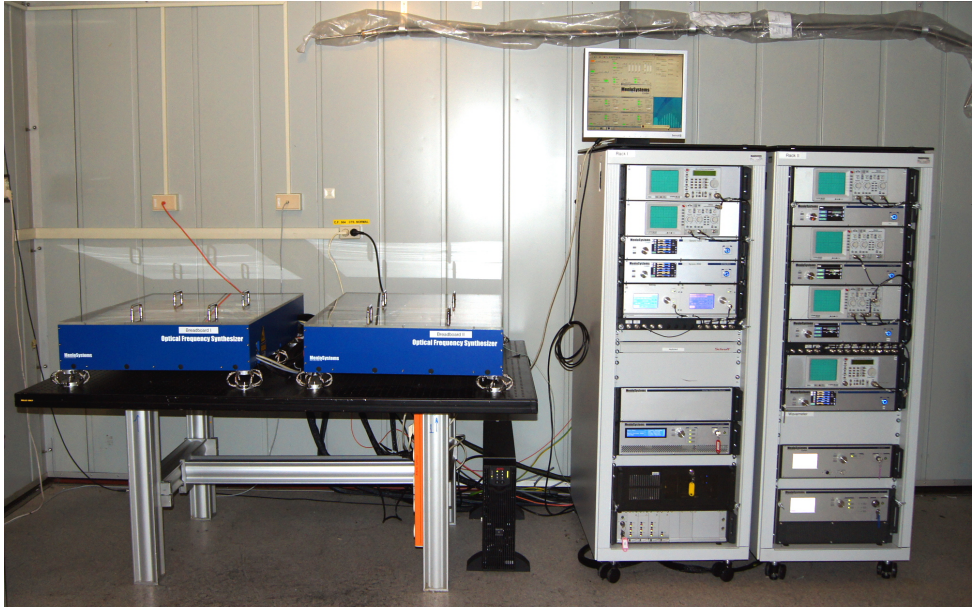

**Figure S1 | Photo of the laser frequency comb (LFC1) after its permanent installation on HARPS.** The optical setup is contained in the blue enclosures on the left-hand-side, while the racks on the right accommodate control electronics and powers supplies, as well as a continuous wave laser, a wavemeter, and diode lasers for pumping of the power amplifier see Fig. 1 of the main article).

*1.2 The laser frequency combs of the 2015 campaign*

In May 2015, HARPS was permanently equipped with an LFC that is presently being prepared to become its new routine calibrator. During the installation run, the permanently deployed LFC (LFC1) was characterized relative to a second, temporarily installed LFC (LFC2), which was shipped back to Germany after the test. In October 2016, LFC2 arrived at the Wendelstein Observatory – its final site of operation – as the calibration system for the recently upgraded FOCES spectrograph[15].

Given the spectral resolution of HARPS of 115 000, an optical resolution element has a 5 GHz full-width at half-maximum (FWHM) in the spectrograph's mid-spectral range. The FWHM being sampled by 3.3 pixels, the 15 µm width of a single pixel spans 1.5 GHz, or 0.8 km/s. Theoretically, the optimal mode spacing of an astronomical LFC is 3 times the FWHM of an optical resolution element[16]. For HARPS, we have decided for a slightly wider spacing of 18 GHz for LFC1. This keeps the calibration lines well apart with virtually no residual overlap, which facilitates data analysis. The mode spacing of LFC2 is 25 GHz, adapted to the slightly lower resolution of the FOCES spectrograph of 70000. Figure S1 shows a photograph of LFC1 after its permanent installation on the site. The configuration of the LFCs is explained in the main article. A more detailed description is found in[17]. It should however be noted that, contrary to the LFC described in[11], the present systems LFC1 and LFC2 no longer employ a second-harmonic generator (SHG). Earlier versions used the SHG to transfer the infrared (IR) laser spectrum into the green before spectral broadening. This is now omitted, as we have developed tapered photonic crystal fibres (PCFs) that directly broaden the IR spectrum far enough to cover the visible range[18].

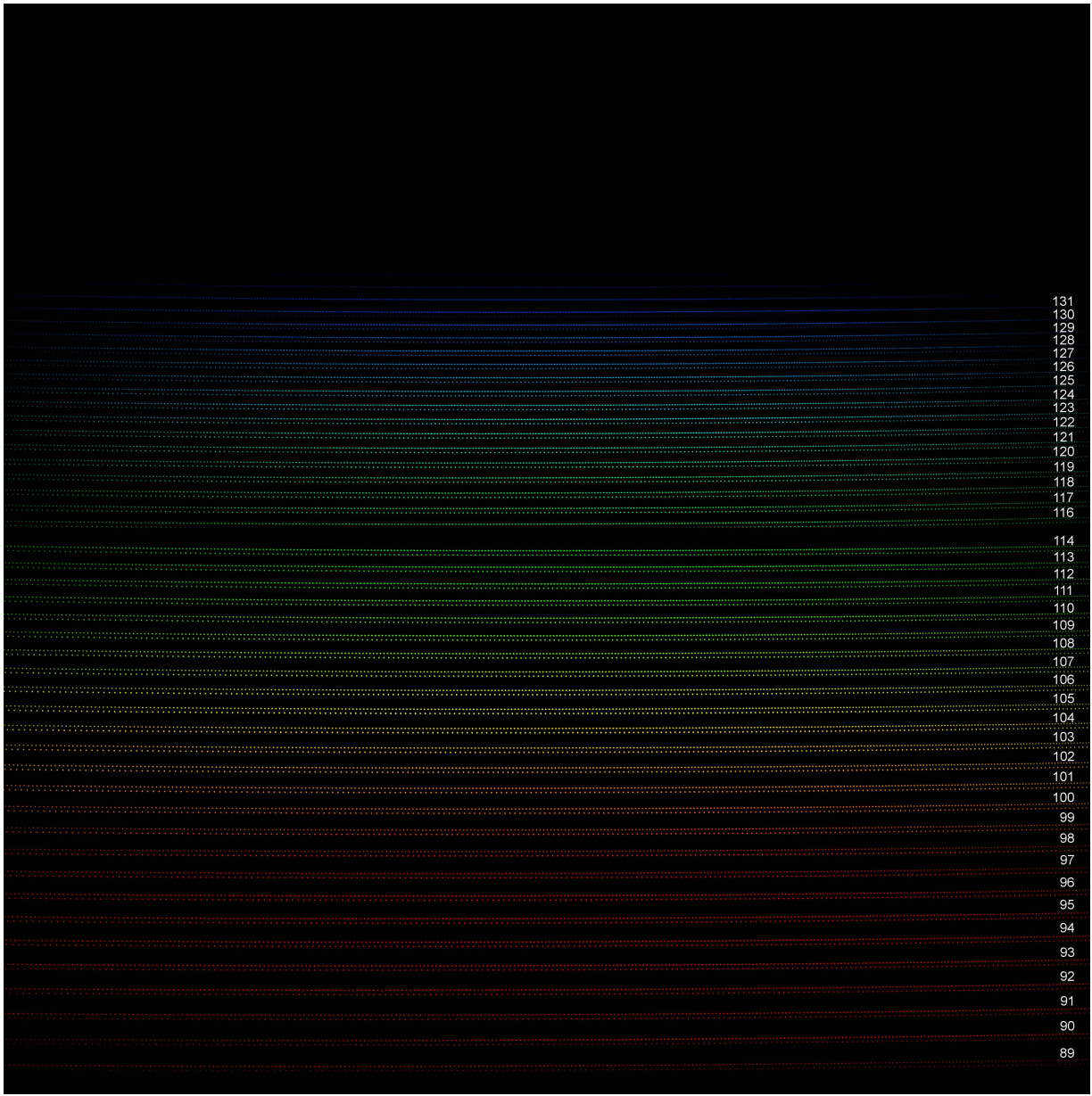

**Figure S2 | Full echellogram with different LFCs on the two channels of HARPS.** The upper half of each order is channel A (LFC1, 18 GHz mode spacing) and the lower part channel B (LFC2, 25 GHz mode spacing). The original grey-scale image as recorded with HARPS has been coloured, using the known wavelengths of the diffraction orders, to approximate the perception of the human eye. The number of the physical diffraction order is annotated above each echelle order.

Figure S2 shows an echellogram with the two LFCs on HARPS. 43 echelle orders are illuminated. The spectrum of LFC1 contains about 12 000 spectral lines over the observed spectral range, many of which appear twice due to the spectral overlap of the echelle orders. The echellogram contains a gap in the green region, where one order of channel A and two orders of channel B are lost. The gap results from the space between the active areas of the two CCD chips that make up the mosaic camera.

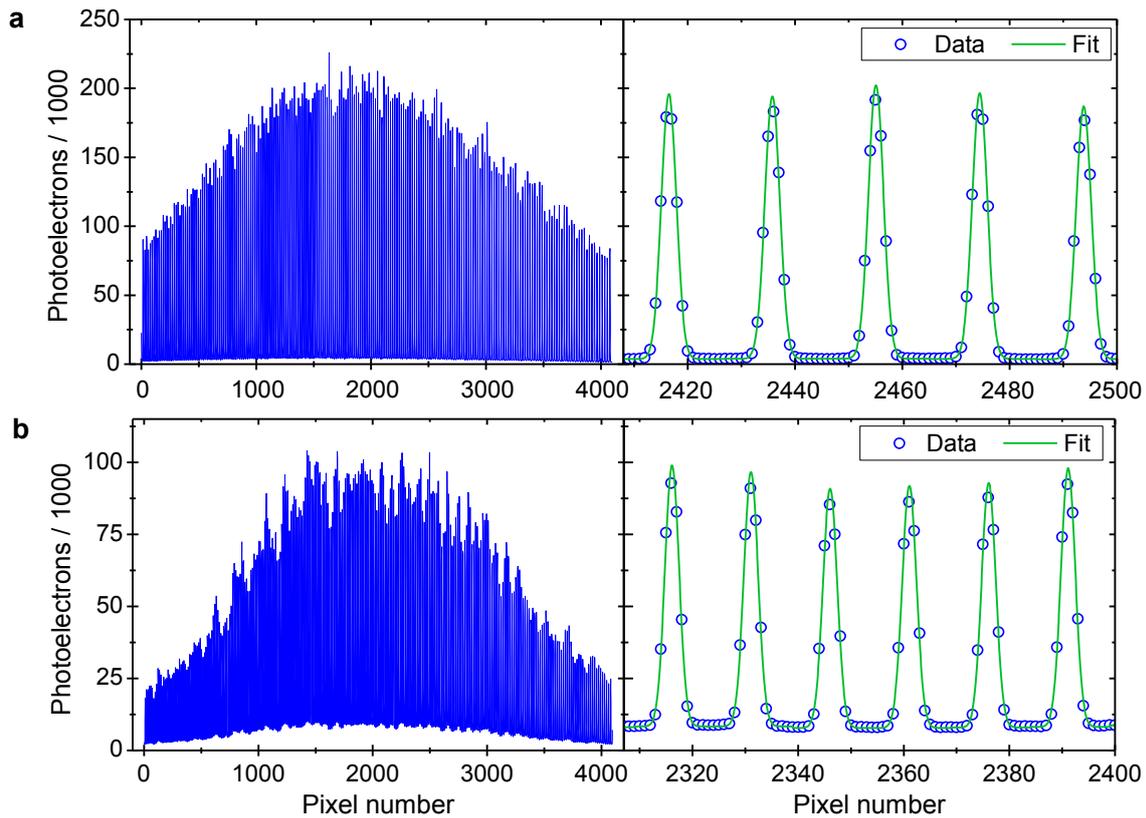

**Figure S3 | Echelle orders of channel B after data extraction, using LFC2 (25 GHz mode spacing).** Left: data over the full order. Right: fit (green) to the data (blue) on a zoomed-in horizontal scale. **a**, Echelle order 100 (centre wavelength 612 nm). **b**, Echelle order 128 (centre wavelength 478 nm). Comparison to part **a** shows a clear trend of the background to rise towards shorter wavelengths.

## 2 Undesired spectral components

### 2.1. Background

In our initial tests of the 2015 campaign, we noticed an unusually strong background component to the LFC spectrum, which was not known from previous campaigns. This is clearly seen in Figure S3, showing how the intensity between the lines forms a plateau instead of dropping to zero. We also observe a clear and consistent tendency of the background to rise towards shorter wavelengths. The only substantial physical difference to earlier versions of our LFC consists in the spectral broadening strategy. Skipping the SHG simplifies the setup, but requires more nonlinear spectral broadening to be driven in the tapered PCF, which is known to create phase noise from amplitude noise. The noise background is very likely to be seeded by amplified spontaneous emission (ASE) from the high-power Yb-fibre amplifier. The amplifier must now generate more optical power, thereby plausibly emitting more ASE.

In fact, this view is confirmed by our later investigations on this matter using HARPS, FOCES, and a home-built echelle spectrograph[19]. Optimizing the core diameter of the gain fibre and its length can reduce the amount of ASE produced, which lowers the background. In case that the ASE spectrum does not fully coincide with the signal of the infrared laser source, it is possible to block a part of the ASE, which again lowers the background. Careful optimization of the amplifier, pre-amplifier and broadening schemes, as well as minimization of losses, should allow us to significantly reduce the background level. For spectrograph calibration, this has the benefit of reducing the photon noise, because the photons forming the background are an additional source of photon noise and decrease the available dynamic range of the CCD[19]. Furthermore, we found that the spectral background can lead to systematic errors, if the background has a slope that is not taken into account by the fit function. With a simple Gaussian fit, we could observe this very clearly on strongly modulated spectra. Since the background level seems to follow the signal strength of the comb lines, the background can acquire a distinct slope through fine modulations in the spectral line intensities that lie below the resolution of the spectral flattening unit. When the spectral structures changed, we observed an apparent global calibration shift of up to 1 m/s. The effect was fully eliminated when adding a first-order polynomial to the fit function (see Methods).

*2.2 Crosstalk and stray light*

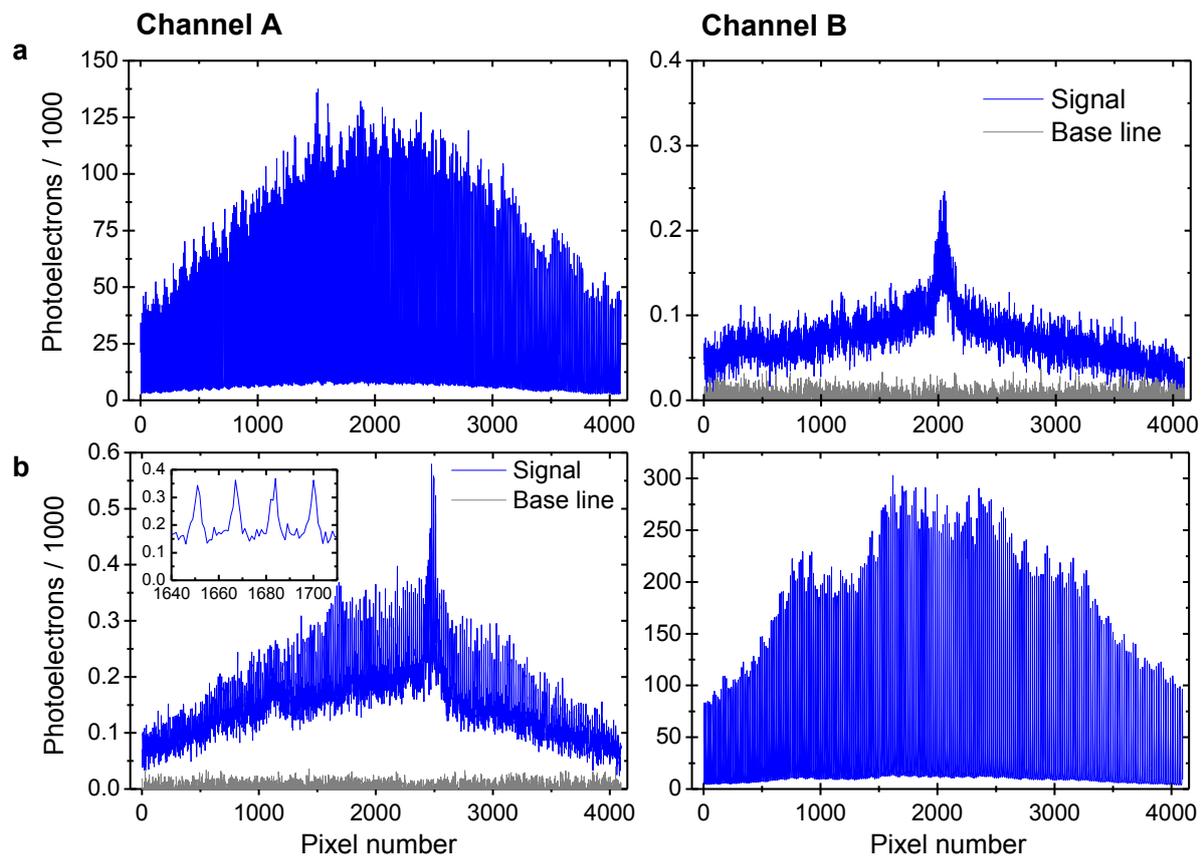

**Figure S4 | Crosstalk characterization.** LFC1 is in channel A (left-hand side) and LFC2 in channel B (right-hand side). For measuring the crosstalk in each direction, an LFC is injected into just one channel, while observing the signal on the other one. The base line (grey) is a measurement without input light on either channel. We deliberately show echelle orders that contain strong ghosting (71 % relative to the strongest ghost found in channel A; 63 % for channel B). **a**, Crosstalk from A to B in order 108 (centre wavelength: 567 nm). **b**, Crosstalk from B to A in order 111 (centre wavelength 551 nm). The inset shows that a comb spectrum is observed in channel A. Notice that in these measurements, LFC2 is more intense than LFC1.

A minor contribution to the spectral background stems from stray light within the spectrograph, which originates mainly from imperfections of the grating. We characterize this contribution in the course of a more general examination of the crosstalk between the two channels. This is done by injecting light into one channel, and observing the signal on the other. Figure S4a shows the crosstalk from channel A to channel B. It consists of two components: stray light and ghosting. Ghosted diffraction orders cross the echelle orders at specific points, creating small artefacts in the spectrum. At their peak, this creates a crosstalk of $1.3 \times 10^{-3}$. The affected sections can be excluded from the analysis, which for us has so far never made a difference. The stray light component amounts to $8 \times 10^{-4}$, and appears to be structureless. Hence, it should not have an influence on spectrograph calibration.

As expected, the crosstalk in opposite direction (see Figure S4b) exhibits ghosting and stray light of the same level. However, contrary to the crosstalk from channel A to B, we here also find a third component that consists in spurious comb lines throughout the entire echelle order. This might either be explained through light pollution in the fibre feed or through charge spilling on the CCD. The effect creates a crosstalk of $5 \times 10^{-4}$. For the LFC this might cause a line shift of about 1 m/s in a worst-case scenario. The combination of an 18 GHz LFC with a 25 GHz LFC leads to recurring line constellations every 450 GHz. The effect is therefore not necessarily fully diluted when averaged over many lines. However, the effect should vanish if two LFCs with the same offset frequency and mode spacing were used, as the signal would only be polluted by another signal with identical lines. Hence, measurements using the same LFC is in both channels should not be impacted. The crosstalk can, however, at least partly be responsible for the absolute disagreement between the two different LFCs (Fig. 5 of the main article).

## 3 Optimization of fibre coupling and scrambling

For each channel, the light is fed through a series of three different multimode fibres before reaching the spectrograph (see Fig. 3 of the main article). The light of the LFC has high spatial coherence, as opposed to that of a thorium-argon lamp. Dynamic fibre scrambling is, therefore, of vital importance to reach a calibration that is stable within the photon noise limit. This scrambles the relative phase of the spatial modes propagating within the fibre, which blurs out their interference pattern ("laser speckles") at the fibre output. We achieve this through agitation of the first and last fibre of the series. The shaker on the last fibre, however, was initially not used, because it is not foreseen for standard operations of the telescope and spectrograph. Instead, we attempted to fully rely on agitating the first fibre only. This requires special care, because in a sequence of several fibres, it is not guaranteed, that the previously imposed relative dephasing of spatial modes is fully passed on to the next fibre(s). Depending on how several cascaded fibres are coupled to one another, some spatial modes of the later fibres might partly share a stable relative phase. The fewer spatial modes of a scrambled fibre are coupled to a non-scrambled fibre, the more the effect of the scrambling will be lost.

Finding a configuration that can best preserve the spatial mode scrambling was an important first goal of our campaign. This meant to optimize the alignment of the fibre illumination in the calibration unit. The optimization was first done with a single LFC on both channels (first with LFC1, then LFC2) before we moved on to testing the relative stability of the two LFCs. We characterized the performance of each configuration in sequences of exposures. The light injection into the fiber was optimized so as to minimize the standard deviation of the differential calibration shifts (A–B) in the respective sequence. Figure S5 shows all exposures obtained in this way with LFC1 in both channels. The overall standard deviation in A–B is 32.1 cm/s, while the average photon noise is 4.0 cm/s. The plot does not only reveal how some configurations provided better stability than others, but it also shows how a change in the alignment systematically shifts the calibration in A–B. Changing the alignment also altered the throughput, which can in part explain the calibration errors. This is because charge-transfer inefficiency (CTI) of the CCD causes line shifts depending on signal strength[20,21]. However, although the calibration shifts and the changes in signal strength seem to have a common cause, they do not relate to each other in a simple and stable way. Trying to adjust the calibration shifts by assuming some simple relationship with signal strength (linear, inverse, exponential, or logarithmic) could reduce the standard deviation in A–B to no less than 29.4 cm/s. This suggests that the varying spatial mode occupation in the fibres is the dominant reason for the shifts, not the altered transmission. Both influences are side-effects of the changing alignment.

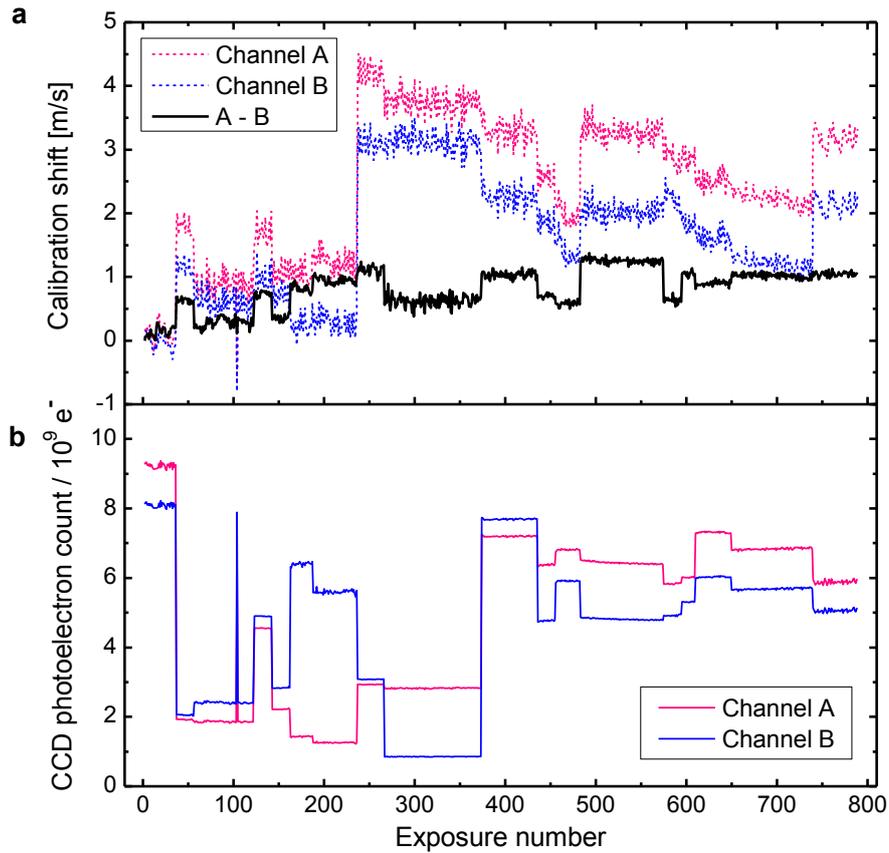

**Figure S5 | All spectrograph exposures with LFC1 on both HARPS channels.** The graph concatenates various series of exposures with differing illumination of the multimode fibres. **a**, Calibration shifts relative to a common reference. As usual, calibration errors are to be seen from the differential shifts A–B, since the channels A and B by themselves are influenced by spectrograph drifts. **b**, Total number of photoelectrons detected in each channel. The offset frequency is 5.95 GHz up to exposure 435, then 5.80 GHz until exposure 574, and finally 5.70 GHz for the remaining part of the data. Mode spacing: 18 GHz.

Although LFC1 was operated at three different offset frequencies, this is taken into account in Figure S5 and can by no means explain the shifts in A–B. Moreover, the interpolation method that we use to relate LFC spectra with different parameters has been tested in an earlier campaign and shown to have no measurable impact on calibration down to at least 7 cm/s[22].

After having optimized the fibre injection in the calibration unit as explained above, we achieved nearly photon noise-limited results with both LFCs. A typical series with LFC1 is show in Figure S6. The standard deviation of the relative shifts is of 4.3 cm/s, which is 34 % above the photon noise of 3.2 cm/s. The excess noise is probably caused by some residual fibre modal noise through less than optimal spatial mode scrambling. When proceeding to measuring the relative stability of the two LFCs, we were first facing a significantly higher excess noise, resulting in a standard deviation of up to 13.3 cm/s with a photon noise of 3.2 cm/s. We then installed a dynamic scrambler to shake the last fibre in the sequence for better suppression of modal noise. With this enhanced scrambling configuration we achieved nearly photon-noise limited results also in the relative measurements (Fig. 4 of the main article).

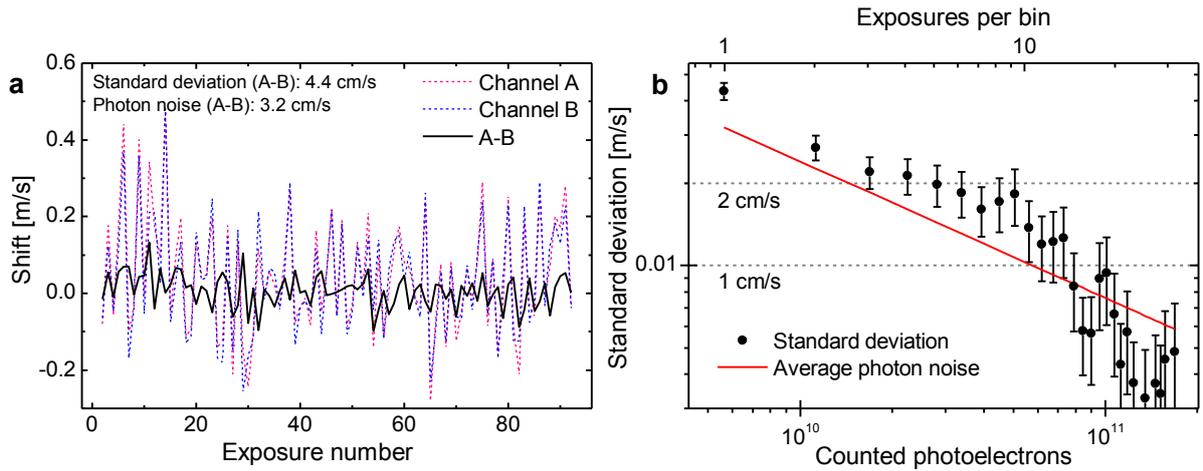

**Figure S6 | Calibration repeatability with LFC1 in both channels A and B.** The alignment of the multimode fibre feed is kept stable during the complete series of exposures. The derived calibration shifts use exposure number 1 as a reference. The series is contained in Figure S5 as running from exposure number 483 to 573. **a**, Calibration shifts without binning of exposures. **b**, Combining several subsequent exposures into binned exposures. The standard deviation in A−B is plotted as a function of the number of exposures per bin. One exposure was triggered every 92 s. Integration time: 40 s. Readout time: 22.6 s.

Later tests with LFC1 permanently installed on HARPS showed that the additional fibre shaker can also be beneficial for closely approaching the photon-noise limit with just one LFC in both channels. However, efforts are still ongoing to make the added shaker obsolete through an optimized coupling of spatial modes in the fibre sequence.

## 4 Absolute wavelength calibration in an astronomical context

Several science cases rely on the unprecedented wavelength accuracy and stability provided by LFCs. It is therefore important to quantify the agreement between the two absolute calibrators at our disposal. Our analysis in the main article reveals a 49 cm/s shift between the two LFCs when considering the average shift in the positions of comb lines. However, astronomical data analysis normally uses a more complex approach, which comprises calibrating scientific exposures by establishing an absolute wavelength scale. To reach the highest precision, this is usually complemented by monitoring line shifts on a second fibre channel. This tracks global shifts in the calibration caused by spectrograph drifts, which was the main focus in the previous parts of this paper. For absolute calibration of echelle spectra, the standard method entails finding a polynomial function relating the positions of a set of emission lines to their known wavelengths. This effectively assigns a wavelength value to each pixel in every echelle order. We emulate the standard astronomical procedure to assess the average agreement of the two LFCs in the important aspect of absolute calibration.

We determine the positions of the lines in the way described in the Methods section and the wavelengths through the means of the LFC equation. The wavelength calibration is derived by fitting an eight order polynomial to this data. We compute the unweighted average shift in the wavelengths of all pixels relative to the reference exposure to quantify the overall shift in absolute wavelength calibration (Figure S7). This approach seamlessly handles the change to a different LFC during a series of exposures. The results obtained are very similar to the results in the main article, with the change to another LFC causing a shift of 53 cm/s (as compared to 49 cm/s with the other method). The results are slightly noisier because the contributions of the lines are not weighted by their photon noise. The analysis is also more prone to the systematic influence of CTI, as it affects weak lines more heavily than strong lines[20,21], which all receive the same weight. Differences in line intensities of the two LFCs may thus explain the 4 cm/s larger disagreement between the two LFCs as determined with this method.

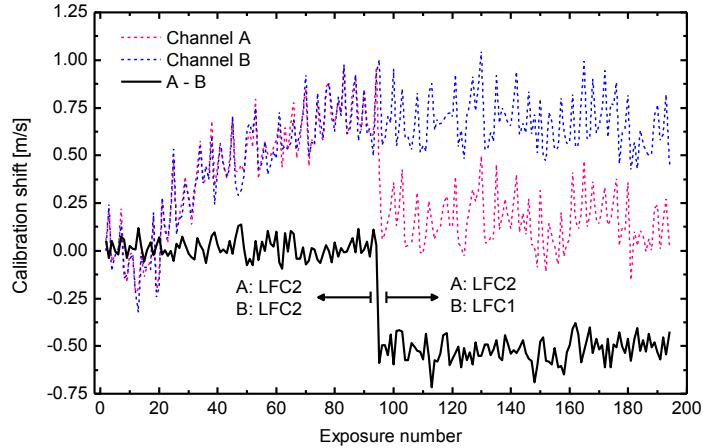

**Figure S7 | Calibration reproducibility with two different LFCs: mean shift in calibration polynomial.** We supplement the analysis of calibration reproducibility in the main article by applying the standard astronomical calibration method of fitting polynomials to calibrate the spectrograph in absolute terms. We find an average shift of 53 cm/s between the calibrations performed by the two LFCs on the same dataset as in Fig. 5. The slightly larger offset using this method is probably caused by altered sensitivity to systematic effects.

## 5 All LFC vs. LFC exposures ever recorded with HARPS

Considering all campaigns that we have so far carried out with LFCs on HARPS, the available data now span a total of more than six years. Although made up from measurements with very different LFCs using different mode spacing and offset frequency, such a long time series is of great value for judging the long-term stability of HARPS and for identifying systematic influences. These aspects are essential when observing phenomena that evolve on extended time horizons such as long-period exoplanets.

Figure S8 shows all HARPS exposures ever recorded with LFC light in both channels. All exposures are related to a common reference of the year 2012. We use linear interpolation between the comb lines of the reference to relate LFCs with different structures (see Methods). While A and B experience a drift by a total of about 35 m/s, A–B exhibits no clear continuous drift. Instead, A–B seems to be dominated by effects from varying illumination conditions. This is most clearly seen in a section of the data that is shaded in grey in Figure S8. Here, one or both channels were attenuated by up to a factor of 100 through grey filters with different optical density. This causes shifts on the m/s scale, which can be explained through CTI of the CCD readout mechanism[20,21] and through effects from the data extraction algorithm of the HARPS pipeline that depend on signal strength.

The systematic errors evident from Figure S8 are of up to several m/s. For measurements requiring cm/s precision, this highlights the importance of minimizing variations in signal strength and changes in fibre-coupling conditions. Future projects should, therefore, place particular emphasis on solutions for enhanced mode scrambling, modelling of CTI, and distortion-free data extraction techniques.

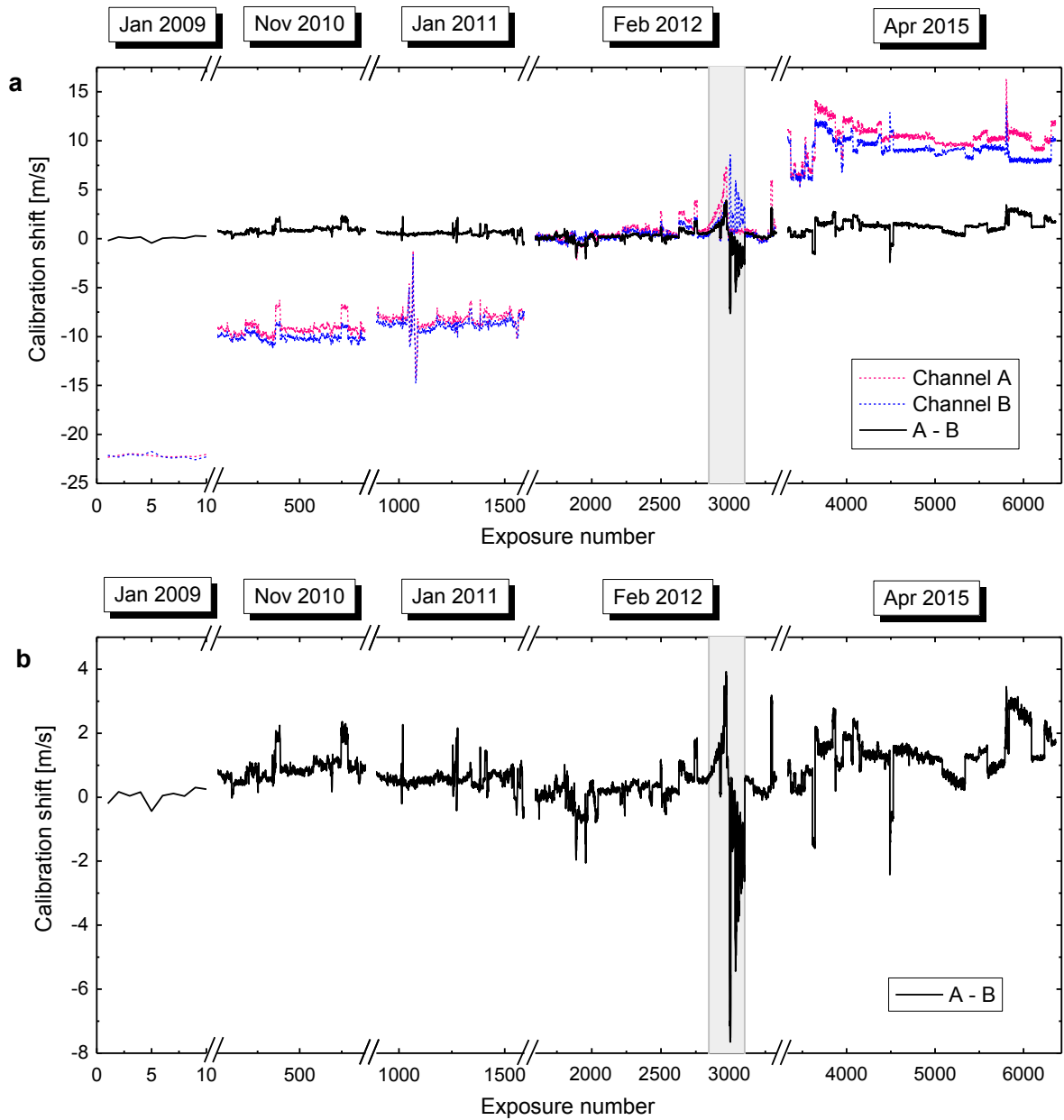

**Figure S8 | All HARPS exposures with LFCs on both channels ever recorded. a**, Evolution of the channels A, B and their difference (A–B). **b**, A–B only, on a magnified vertical scale. In the grey-shaded region, one or both channels are attenuated by neutral-density filters of varying optical density. The reference is an exposure recorded in the year 2012.

## References

1. Mayor, M. *et al.* Setting New Standards with HARPS. *The Messenger* **114,** 20–24 (2003).

2. Rupprecht, G. *et al.* The exoplanet hunter HARPS: performance and first results. *Proc. SPIE* **5492,** 148–159 (2004).

3. Wilken, T. *et al.* High-precision calibration of spectrographs. *Mon. Not. R. Astron. Soc.* **405,** L16–L20 (2010).

4. Wilken, T. *et al.* A spectrograph for exoplanet observations calibrated at the centimetre-per-second level. *Nature* **485,** 611–614 (2012).